\documentclass[11pt,a4paper]{article}
\usepackage{amstex,amsthm,amssymb,epsfig}
\usepackage{cite,indentfirst}

\def\C|{{\mathbb C} \,}
\def\B|{{\mathbb B} \,}
\def\S|{{\mathbb S} \,}
\def\G|{{\mathbb G} \,}
\def\N|{{\mathbb N} \,}
\def\F|{{\mathbb F} \,}
\def\K|{{\mathbb K} \,}

\def\cM{\mathcal M}
\def\cN{\mathcal N}
\def\cV{\mathcal V}
\def\cS{\mathcal S}
\def\cQ{\mathcal Q}
\def\cX{\mathcal X}

\def\det{\operatorname{det}}

\def\tensor{\otimes}

\newcommand{\bra}[1]{\langle\,#1\,|}
\newcommand{\ket}[1]{|\,#1\,\rangle}

\def\partialsur#1{\frac{\partial}{\partial#1}}

\newtheorem{prop}{Proposition}[section]

   {
   }

\renewcommand{\theequation}{\thesection.\arabic{equation}}

\def\beqa{\begin{eqnarray}}
\def\eeqa{\end{eqnarray}}
\def\ba{\begin{array}}
\def\ea{\end{array}}

\def\l{\langle}
\def\a{\alpha}

\def\la{\lambda}
\def\s{\sigma}
\def\sul{\sum\limits}
\def\pl{\prod\limits}
\def\lt({\left(}
\def\rt){\right)}

\def\prodmu #1{\prod\limits_{j #1 k} \sin(\mu_k-\mu_j)}
\def\prodla #1{\prod\limits_{j #1 k} \sin(\lambda_k-\lambda_j)}

\begin{document}
\begin{titlepage}
\begin{flushright}
LPENSL-TH-13/98\\
\end{flushright}
\par \vskip .1in \noindent

\begin{center}
{\LARGE Spontaneous magnetization of the XXZ Heisenberg spin-$ \frac 12$  chain}\\
\end{center}
  \par \vskip .3in \noindent

\begin{center}

      {\bf A.G. IZERGIN$^{*}$, N. KITANINE$^{*}$,  J. M. MAILLET, V. TERRAS}
  \par \vskip .1in \noindent

{\sl  Laboratoire de Physique $^{**}$\\
Groupe de Physique Th\'eorique\\
       ENS Lyon, 46 all\'ee d'Italie 69364 Lyon CEDEX 07
       France}\\[0.6in]
\end{center}

\par \vskip .10in
\begin{center}
{\bf Abstract}\\
\end{center}

\begin{quote}
Determinant representations of form factors
are used to represent
the spontaneous magnetization of the Heisenberg XXZ chain ($\Delta >1$) 
on the finite
lattice as the ratio of two determinants. In the thermodynamic limit 
(the lattice of infinite length), the Baxter formula is reproduced
in the framework of Algebraic Bethe Ansatz. It is
shown that the finite size corrections to the Baxter formula are exponentially 
small. 
\end{quote}
\par \vskip .5in

\begin{flushleft}
\rule{5.1 in}{.007 in}\\
$^{*}$ {\small Permanent address: St Petersburg Department
 of the Steklov Mathematical Institute, Fontanka 27, St Petersburg 191011, Russia.}\\
$^{**}${\small URA 1325 du CNRS, associ\'ee \`a  l'Ecole
Normale Sup\'erieure de Lyon.}\\
{\small This work is supported by CNRS (France), the EC-TMR contract FMRX-CT96-0012},  MENRT (France) fellowship AC 97-2-00119 and MAE (France) fellowship 96/9804.\\
{\small email: izergin\symbol{'100}pdmi.ras.ru, nkitanin\symbol{'100}enslapp.ens-lyon.fr, maillet\symbol{'100}enslapp.ens-lyon.fr, vterras\symbol{'100}enslapp.ens-lyon.fr}\\[0.2 in]

December 1998
\end{flushleft}

\end{titlepage}

%%%%%%%%%%%%%%%%%%%%%%%%%%%%%%%%%%%%%%%%%%%%%%%%%%%%%%%%%%%%%%%%%%%%%%%%%%%%%%%%
\section{Introduction}
\setcounter{equation}{0}

The aim of this paper is the computation, within algebraic Bethe ansatz method,
of the spontaneous magnetization of the XXZ spin-$\frac 12$ chain
(see \cite{Bax1, Bax2}),
by taking the thermodynamic limit of the form factor formulas of local spin
operators for the finite chain obtained in \cite{kmt}.

\bigskip

We consider the XXZ Heisenberg chain  of finite length $M$ \cite{Heis,Beth},
the Hamiltonian of which is given by
\begin{equation}
\label{ham}
  H_\Delta=J\sum_{m=1}^M \Big\{ \sigma^x_m \sigma^x_{m+1} +
  \sigma^y_m\sigma^y_{m+1} + \Delta(\sigma^z_m\sigma^z_{m+1}-1)\Big\},
\end{equation}
with periodic boundary conditions
\[\s^a_1=\s^a_{M+1},\quad a=x,y,z.\]
This Hamiltonian acts in the quantum space $\mathcal{H}$
of the chain, which is the tensor product
of $M$ local quantum spin-$\frac{1}{2}$ spaces $\mathcal{H}_m$
isomorphic to $\mathbb{C}^2$ and $\s^a_m,\ a=x,y,z,$ are the standard Pauli matrices 
acting in $\mathcal{H}_m$. In the anisotropic case ($\Delta
\neq \pm 1 $) this model was solved by means of the Bethe ansatz in
\cite{YY1,YY2} (see also \cite{g}).

We use the algebraic Bethe ansatz solution for this model \cite{ft}, taking
the XXZ $R$-matrix in the following normalization
(as a matrix acting in $\mathbb{C}^2 \tensor \mathbb{C}^2$):
\begin{equation}
   R(\la,\mu)=
  \left(\ba{cccc}1&0&0&0\\
                 0&b(\la,\mu)&c(\la,\mu)&0\\
                 0&c(\la,\mu)&b(\la,\mu)&0\\
                 0&0&0&1\ea\right).
\end{equation}
The functions $b(\la,\mu)$ and $c(\la,\mu)$ are defined as
\[b(\la,\mu)=\frac{\sin(\la-\mu)}{\sin(\la-\mu-i\zeta)},\quad c(\la,\mu)
=-\frac{i\sinh\zeta}{\sin(\la-\mu-i\zeta)},\]
where the parameter $\zeta$ is related to the anisotropy parameter
$\Delta$ of the Hamiltonian as
\begin{equation*}
  \Delta=\frac{1}{2}(q+q^{-1}),\quad \text{with}\ q=e^\zeta.
\end{equation*}
The $R$-matrix is a linear operator in the tensor product of two
two-dimensional
linear spaces $V_1 \otimes V_2$, where each $V_i$ is isomorphic to
${\mathbb C}^2$, and depends generically on two spectral parameters $\lambda_1$ and $\lambda_2$ associated to these two vector spaces. It is denoted by $R_{12} (\lambda_1, \lambda_2)$.

The monodromy matrix is constructed as an ordered product, in an auxiliary
space $V_0$ isomorphic to $\mathbb{C}^2$, of such
$R$-matrices $R_{0 m}$ acting in $V_0 \tensor \mathcal{H}_m$:
\[T(\la)=R_{0 M}(\la+i\frac \zeta2)\dots
           R_{0 2}(\la+i\frac \zeta2)R_{0 1}(\la+i\frac \zeta2)
        =\left(\ba{cc}
                    A(\la)& B(\la)\\
                    C(\la)& D(\la)\ea\right)_{[0]}.\]
In the last formula, the monodromy matrix is represented as a $2\times 2$
matrix
in the auxiliary space $V_0$, whose entries
$A(\la)$, $B(\la)$, $C(\la)$, and $D(\la)$ are operators
in the quantum space $\mathcal{H}$ of the chain.

Since we will further use some results of \cite{kmt}, let us note, to
compare notations, that the
spectral parameters are equal here to the corresponding spectral
parameters of \cite{kmt} multiplied by $i$, and that $\zeta$ corresponds to
$-\eta$ of \cite{kmt}. Moreover, we consider here the homogeneous case
where all the inhomogeneity parameters $\xi_m$ of \cite{kmt} are equal to
$-\zeta/2$.

\bigskip

The eigenstates of the Hamiltonian  (\ref{ham}) can be constructed by the
action
of the operators $B(\la)$ on the ferromagnetic state $\ket{0}$ (which is the
state
with all the spins up). More precisely, the state
$B(\la_1)\dots B(\la_N)\ket{0}$
is an eigenstate of the Hamiltonian \eqref{ham} if the set of spectral
parameters $\{\la_j\}_{1 \le j \le N}$ is a solution of the Bethe equations
\begin{equation}
  \left(\frac{\sin(\la_j+\frac {i\zeta}2)}{\sin(\la_j-\frac
{i\zeta}2)}\right)^M
         \pl_{k=1}^N\frac{\sin (\la_j-\la_k-i\zeta)}
                         {\sin (\la_j-\la_k+i\zeta)}=-1,
  \quad 1 \le j \le N.
\label{Bethe1}
\end{equation}

The norm of Bethe eigenstates  is given by the Gaudin formula \cite{g,Kor}
(see also \cite{kmt}):
\begin{equation}
  \bra{0}\,
%\textstyle{
\pl_{j=1}^N
%}
C (\la_j) \,
%\textstyle{
\pl_{k=1}^N
%}
B (\la_k)\, \ket{0}\
        = (-\sinh \zeta)^N
                \pl_{\alpha\ne\beta}
           \frac{\sin(\la_\alpha- \la_\beta+i\zeta)}
                {\sin(\la_\alpha- \la_\beta)}
            \det \cN (\{\la_\a\}),
\label{gaudin}
\end{equation}
where $\cN$ is an $N \times N$ matrix the elements of which are given by
\begin{align}
  \cN_{ab} &= -i\partialsur\la_b
                       \ln\bigg\{
           \biggl(\frac{\sin(\la_a-\frac {i\zeta}2)}
                       {\sin(\la_a+\frac {i\zeta}2)}\biggr)^M
        \prod_{k=1 \atop k\ne a}^N
                \frac{\sin(\la_a- \la_k+i\zeta)}{\sin(\la_a- \la_k-i\zeta)}
                       \bigg\}.
%\\
%                    &= \delta_{ab}\biggl(-\partialsur\mu_a \ln r(\mu_a)
%                          +\sum_{\alpha=1}^n
%                             \frac{2\eta}{(\mu_a-\mu_\alpha)^2-\eta^2}\biggr)
%                        -\frac{2\eta}{(\mu_a-\mu_b)^2-\eta^2}.
\label{matrix-gaudin}
\end{align}

The ground state of the XXZ model in the region $\Delta=\frac 12(q+q^{-1})>1$,
$q>1,$
 is degenerated in the thermodynamic limit ($M\rightarrow\infty$),
namely there are two
states with the same energy which we will call the
ground state $\ket{\Psi_1}$ and the quasi-ground state
$\ket{\Psi_2}$ (on the finite lattice, these states possess
different energies). These two
states  can be obtained in the algebraic Bethe ansatz framework (see \cite{g}).

The spontaneous  magnetization
is defined as the modulus of the normalized
matrix element of the local spin operator $\s_m^z$ between these two states:
\begin{equation}
\label{SM}
  s_0=\left|\frac {\bra{\Psi_1}\,\s_m^z\,\ket{\Psi_2}}
                  {\l\Psi_1\ket{\Psi_1}^{\frac 12}
         \l\Psi_2\ket{\Psi_2}^{\frac 12}}\right|.
\end{equation}
In \cite{Bax2,BaxCTM1}, Baxter proved that the spontaneous magnetization of the
infinite chain could be expressed as an infinite product:
\begin{equation}
  s_0=\left(\pl_{n=1}^\infty\frac{1-q^{-2n}}{1+q^{-2n}}\right)^2,\qquad
    (M=\infty).
\label{Baxter}
\end{equation}
This formula in the thermodynamic limit was also reproduced by means
of the q-vertex operator approach (see \cite{JM1}).

In this paper, we study the spontaneous magnetization $s_0$
in the framework of the algebraic Bethe ansatz. From the results of
paper \cite{kmt}, the corresponding form factor on the finite chain
is represented as a ratio of two
determinants of
$M\times M$ matrices. In the thermodynamic limit, it becomes the ratio of
two Fredholm determinants of linear integral operators, which can be
computed explicitly. This leads directly to
the Baxter formula (\ref{Baxter}).
Thus, we obtain here an independent proof of the Baxter formula
from algebraic Bethe ansatz.
We prove also
that there are no perturbative ($\frac 1{M^k}$)
finite size corrections so that for the finite chain
the Baxter formula has only exponentially small  corrections.

\bigskip

This paper is organized as follows.

In section \ref{sec:formfac}, the expression for the
form factor of operator $\s^z_m$ obtained in \cite{kmt} is recalled. It is
reformulated in a form suitable for the thermodynamic limit and particularized
to the ground and quasi-ground states.

In section \ref{sec:thermlim}, the procedure of taking  the thermodynamic limit
is discussed.

In section \ref{sec:result}, the spontaneous magnetization is represented
in terms of Fredholm determinants, which gives a proof of the
Baxter formula in the thermodynamic limit. The
finite size corrections are also evaluated.

%%%%%%%%%%%%%%%%%%%%%%%%%%%%%%%%%%%%%%%%%%%%%%%%%%%%%%%%%%%%%%%%%%%%%%%%%%%%%%%
\section{The form factor\label{sec:formfac}}
\setcounter{equation}{0}

Let us consider the function
\begin{equation}
  s_z(m)=
    \frac{\bra{0}\pl_{j=1}^{N}C(\mu_j)\,\s_m^z\,\pl_{k=1}^{N}B(\la_k)\ket{0}}
         {\bra{0}\pl_{j=1}^{N}C(\la_j)\,\pl_{k=1}^{N}B(\la_k)\ket{0}},
\label{start}
\end{equation}
where $\{\la_k\}_{1\le k \le N}$ and $\{\mu_j\}_{1\le j \le N}$ are solutions
of the Bethe equations
\begin{equation}
  \left(\frac{\sin(\la_j+\frac {i\zeta}2)}{\sin(\la_j-\frac
{i\zeta}2)}\right)^M
         \pl_{k=1}^N
      \frac{\sin (\la_j-\la_k-i\zeta)}{\sin (\la_j-\la_k+i\zeta)}=-1,
    \quad 1 \le j \le N.
\label{Bethe}
\end{equation}
 These equations can also be written in a logarithmic form:
\begin{equation}
     M p_0(\la_j)+\sul_{k=1}^N\theta(\la_j-\la_k)=2\pi n_j,
     \quad 1\le j \le N,
\label{bethelog}
\end{equation}
where $n_j$ are integers for $N$ odd and half integers for $N$ even. The bare
momentum $p_0(\la)$ and the scattering phase $\theta(\la)$ are defined as
\begin{align*}
  p_0(\la) &=i\ln\frac{\sin(\la+\frac {i\zeta}2)}{\sin(\la-\frac {i\zeta}2)},\\
  \theta(\la)&=i\ln\frac{\sin(i\zeta-\la)}{\sin(i\zeta+\la)}.
\end{align*}

\bigskip

In the recent paper \cite{kmt},  an expression for any arbitrary
matrix element (form factor) of the $\s_m^z$ operator between
two Bethe states has been obtained in the framework of algebraic Bethe ansatz.
Using this result (see proposition 5.2 of \cite{kmt})
as well as the Gaudin-Korepin \cite{g,Kor} formula for the
norm of Bethe states, one can represent the function  $s_z(m)$ as follows:
\begin{equation}
   s_z(m)=\exp\bigg\{-i(m-1)\sul_{k=1}^N (p_0(\la_k)-p_0(\mu_k))\bigg\}
         \prod_{j>k}\frac{\sin(\la_k-\la_j)}{\sin(\mu_k-\mu_j)}
             \frac{\det(\cS-2\cQ)}{\det \cN},
\label{formfactor}
\end{equation}
where the $N\times N$ matrix $\cS$ is given by
\begin{gather*}
   \cS_{a b}=\Phi_{a b}^+ +\Phi_{a b}^-,\\
   \Phi^\pm_{a b}=
       \frac{\sinh\zeta}{\sin(\la_b-\mu_a)\sin(\la_b-\mu_a\pm i\zeta)}
       \pl_{k=1}^N\frac{\sin(\la_b-\mu_k\pm i\zeta)}
                       {\sin(\la_b-\la_k\pm i\zeta)},
\end{gather*}
and $\cQ$ is the following $N \times N$ rank one matrix:
\[
  \cQ_{a b}=\frac{\sinh\zeta}{\sin(\mu_a+i\frac\zeta 2)
\sin(\mu_a-i\frac\zeta 2)}
           \pl_{k=1}^N\frac{\sin(\mu_k-i\frac\zeta 2)}
                           {\sin(\la_k-i\frac\zeta 2)}.
\]
$\cN$ is the Gaudin matrix:
\begin{equation}\label{mat:N}
  \cN_{a b}=\delta_{a b}
            \big\{ M p_0'(\la_a)-2\pi\sul_{k=1}^N K(\la_a-\la_k)\big\}+
                              2\pi K(\la_a-\la_b),
\end{equation}
where the function $K(\la)$ is proportional to the derivative of the
scattering phase
\begin{equation}
  K(\la)=-\frac 1{2\pi}\theta'(\la)=
             \frac{\sinh 2\zeta}{2\pi\sin(\la+i\zeta)\sin(\la-i\zeta)}.
\label{kernelK}
\end{equation}
Note that the fact that $\cQ$ is a rank one matrix together with the orthogonality
of two different Bethe states (i.e. $\det \cS=0$) allowed us to insert the
global factor
$\prod_{k=1}^N \frac{\sin(\mu_k-i\zeta/2)}{\sin(\la_k-i\zeta/2)}$
in the matrix elements of $\cQ$.

\bigskip

Thus, we will derive the Baxter formula from the expression~\eqref{formfactor}.
In order to take the thermodynamic limit in~\eqref{formfactor}, we first
rewrite it in a more convenient form by introducing the following matrix
\begin{equation}
\label{matrixX}
  \cX_{a b}=
    \frac{e^{i(\la_a-\mu_b)}}{\sin(\la_a-\mu_b)}
    \frac{\pl_{k=1}^N\sin(\mu_b-\la_k)}{\pl_{k\neq b}\sin(\mu_b-\mu_k)},
\end{equation}
the determinant of which is
\[
  \det \cX=\exp\bigg\{i\sul_{j=1}^N(\la_j-\mu_j)\bigg\}
                \frac{\prodla >}{\prodmu >}.
\]
Hence \eqref{formfactor} can be rewritten as
\[
  s_z(m)=\exp\bigg\{-i(m-1)\sul_{k=1}^N (p_0(\la_k)-p_0(\mu_k))\bigg\}
            \exp\bigg\{i\sul_{j=1}^N(\mu_j-\la_j)\bigg\}
    \frac{\det(\cX\cS-2\cX\cQ)}{\det \cN}.
\]
The products of matrices $\cM\equiv \cX\cS$ and $\cV\equiv \cX\cQ$ can be
calculated
using some identities for the rational functions.
The procedure is similar to the one used in
Appendix A. Here we merely give the results:
\begin{align}
  \cM_{a b} &=-2\pi K(\la_a-\la_b)+ \nonumber\\
    &+ i \delta_{a b}
    \frac{\pl_{k\neq a}\sin(\la_a-\la_k)}{\pl_{k=1}^N\sin(\la_a-\mu_k)}
    \left(\pl_{k=1}^N\frac{\sin(\la_a-\mu_k+ i\zeta)}
    {\sin(\la_a-\la_k+ i\zeta)}-
      \pl_{k=1}^N\frac{\sin(\la_a-\mu_k- i\zeta)}
                      {\sin(\la_a-\la_k- i\zeta)}\right),\label{mat:M}\\
  \cV_{a b} &=i\biggl(
     \frac{\exp\{i(\la_a-i\frac\zeta 2)\}}{\sin(\la_a-i\frac\zeta 2))}-
     \frac{\exp\{i(\la_a+i\frac\zeta 2)\}}{\sin(\la_a+i\frac\zeta 2))}
     \exp\bigg\{ i\sul_{j=1}^N\bigl(p_0(\mu_j)-p_0(\la_j)\bigr) \bigg\}\biggr).
     \label{mat:V'}
\end{align}

Note that these expressions are valid for any two different solutions
$\{\la_j\}_{1\le j \le N}$ and $\{\mu_k\}_{1\le k \le N}$
of Bethe equations. To calculate the spontaneous magnetization~\eqref{SM},
one has to particularize them to the ground and quasi-ground states
$\ket{\Psi_1}$ and $\ket{\Psi_2}$, that is to compute the quantity
\begin{equation}\label{szm}
   s_z(m)=\frac {\bra{\Psi_1}\,\s_m^z\,\ket{\Psi_2}}{\l\Psi_2\ket{\Psi_2}}.
\end{equation}

  From now on, $\{\mu_1,\dots\mu_N\}$ and $\{\la_1,\dots\la_N\}$ will
denote the sets of spectral parameters solutions of the Bethe
equations~\eqref{bethelog} corresponding respectively
to the ground state $\ket{\Psi_1}$ and the quasi-ground state $\ket{\Psi_2}$.
For both states, $N$ is equal to $M/2$.
The ground state $\ket{\Psi_1}$ is
parametrized by the following set of $n_j$:
\begin{equation}\label{nj}
    n_j=-\frac{N+1}2+j, \qquad j=1,2,\dots,N,
\end{equation}
whereas the quasi-ground state $\ket{\Psi_2}$ is parametrized by the shifted
set:
\begin{equation}\label{ntj}
    \tilde{n}_j=-\frac{N+1}2+j-1, \qquad j=1,2,\dots,N,
\end{equation}
(see, e.g., \cite{g}).
We will moreover use the following notations:
\[\delta\lambda_j=\la_{j+1}-\la_j,\quad \delta\mu_j=\mu_{j+1}-\mu_j,\quad
\delta\tilde{\lambda}_j=\mu_{j}-\la_j.\]
The energies of these two states are equal in the thermodynamic limit, the
difference
of their total momenta being equal to $\pi$.

\bigskip

Thus, one has expressed the quantity~\eqref{szm} in the form:
\begin{equation}
   s_z(m)=(-1)^{m-1}\exp\bigg\{i\sul_{j=1}^N\delta\tilde{\la}_j\bigg\}
                        \frac{\det(\cM-2\cV)}{\det \cN},
\label{turned}
\end{equation}
where $\cM$ and $\cN$ are the $N\times N$ matrices the elements of which are
given by~\eqref{mat:M} and \eqref{mat:N}, whereas the elements of the matrix
$\cV$ are simply (since the difference of the total momenta of the ground
and quasi-ground state is $\pi$):
\begin{equation}
  \cV_{a b}=\frac{e^{2i\la_a}-\cosh\zeta}
               {\sin(\la_a+i\frac \zeta 2)\sin(\la_a-i\frac \zeta 2)}.
\label{mat:V}
\end{equation}
The expression~\eqref{turned} turns out to be
very convenient to take the thermodynamic
limit since the functional form of the non-diagonal terms
of the matrices $\cM$ and $\cN$, as well as
the matrix elements of $\cV$, do not depend on the number of sites $M=2N$ of
the chain.
Nevertheless, the diagonal terms of the matrices $\cM$ and $\cN$ are more
complicated and should be treated separately.

%%%%%%%%%%%%%%%%%%%%%%%%%%%%%%%%%%%%%%%%%%%%%%%%%%%%%%%%%%%%%%%%%%%%%%%%%%%%%%%%
\section{The thermodynamic limit \label{sec:thermlim}}
\setcounter{equation}{0}

%%%%%%%%
\subsection{General procedure}

In the previous section, we obtained an expression of the spontaneous
magnetization as a quotient of two determinants. Most of the matrix elements
involved are independent of $N$, which makes the thermodonamic limit
($M=2N\rightarrow\infty$) quite obvious. The only terms we have to take care
of (i.e. which depend explicitly of $N$) are on the one hand the global
coefficient, and on the other hand the diagonal matrix elements of $\cM$
and $\cN$. Both kinds of terms consist of sums (or products) on the different
spectral parameters in the ground or in the quasi-ground states. Thus, the
first step in taking the thermodynamic limit is to know how to deal with
such sums. This leads us to formulate the two
following propositions.

As previously, $\{\la_j\}_{1\le j\le N}$ and $\{\mu_j\}_{1\le j\le N}$
denote respectively the set of spectral parameters corresponding to the
quasi-ground state, solution of~\eqref{bethelog}-\eqref{ntj}, and the set
of spectral parameters corresponding to the ground state, solution
of~\eqref{bethelog}-\eqref{nj}.

\begin{prop}
  Let $f$ be a $\mathcal{C}^\infty$ $\pi$-periodic function.
  Then the sum of all the values $f(\la_j)$ where the set of
  the spectral parameters $\{\la_j\}_{1\le j \le N}$ parametrizes the
  quasi-ground state can be replaced by an integral in the thermodynamic limit
  according to the following rule:
  \begin{equation}
      \frac 1M \sul_{j=1}^N f(\la_j)=\int_{-\frac \pi 2}^{\frac \pi 2}
                              d\lambda\,f(\la)\rho(\la)+O(M^{-\infty}),
  \label{prop}
  \end{equation}
  where $\rho(\la)$ is defined as the solution of the Lieb equation \cite{Lieb}
  \begin{equation}
      \rho(\la)+\int_{-\pi/2}^{\pi/2}d\nu\, K(\la-\nu)\rho(\nu)
               =\frac{p_0'(\la)}{2\pi}.
  \label{Lieb}
  \end{equation}
  The same equations~\eqref{prop}-\eqref{Lieb} stand concerning the sum on
  the values $f(\mu_j)$ where $\{\mu_j\}_{1\le j \le N}$ parametrizes
  the ground state.
  \label{period}
\end{prop}

The proof of the proposition is given in the Appendix B.

There exists a generalization of this result in the case that $f$
is not itself a periodic
function but that its derivative
is periodic:

\begin{prop}
  Let $f$ be a $\mathcal{C}^\infty$ function such that $f'$ is $\pi$-periodic.
  Then the sums of the values of $f$ at the points
  $\la_j,\ 1\le j\le N$ parametrizing the quasi-ground state, and
  respectively at the points $\mu_j,\ 1\le j\le N$ parametrizing the
  ground state, are given in the thermodynamic limit as the
  corresponding integrals, the only rational finite-size correction
  being of order $\frac{1}{M}$:
  \begin{align}\label{prop2}
     \frac 1M \sul_{j=1}^N f(\la_j)&=
            \int_{-\frac \pi 2}^{\frac \pi 2}d\lambda\,f(\la)\rho(\la)
                        +\frac {c_1(f)}M+O(M^{-\infty}),\\
     \frac 1M \sul_{j=1}^N f(\mu_j)&=
            \int_{-\frac \pi 2}^{\frac \pi 2}d\mu\,f(\mu)\rho(\mu)
                        +\frac {c_2(f)}M+O(M^{-\infty}),
  \end{align}
  where $c_1$ and $c_2$ are constants independent of $M$.
  \label{period'}
\end{prop}

%%%%%%%%
\subsection{The norm}

Using the proposition~\ref{period} and the Lieb equation~\eqref{Lieb}
one can immediately calculate the diagonal terms of the Gaudin matrix
$\cN$:
\begin{equation}
    \cN_{a a}=2\pi M \big\{\rho(\la_a)+\frac 1M  K(0)\big\}+O(M^{-\infty}).
\end{equation}
 Thus the matrix elements of  $\cN$ have the following form in the
thermodynamic limit:
\begin{equation}
    \cN_{a b}=2\pi M \big\{\delta_{a b}\rho(\la_a)+\frac 1M
K(\la_a-\la_b)\big\}
                                 +O(M^{-\infty}).
\label{N}
\end{equation}

%%%%%%%%
\subsection{Diagonal matrix elements of $\cM$}

In order to calculate the diagonal elements of the matrix $\cM$, let us
consider the
following $\pi$-periodic functions:
\begin{align}\label{Phipm}
     \phi_\pm(\la_a) &=\pl_{k=1}^N\frac{\sin(\la_a-\mu_k\pm i\zeta)}
                                       {\sin(\la_a-\la_k\pm i\zeta)},\\
     \phi(\la_a) &=M\frac{\pl_{k=1}^N\sin(\la_a-\mu_k)}
                         {\pl_{k\neq a}\sin(\la_a-\la_k)},
\end{align}
and those defined similarly for the second set of parameters:
\begin{align}\label{Psipm}
    \psi_\pm(\mu_k) &=\pl_{a=1}^N
           \frac{\sin(\mu_k-\la_a\pm i\zeta)}{\sin(\mu_k-\mu_a\pm i\zeta)},\\
    \psi(\mu_k) &=M\frac{\pl_{a=1}^N\sin(\mu_k-\la_a)}
                        {\pl_{a\neq k}\sin(\mu_k-\mu_a)}.
\end{align}
The diagonal elements of the matrix $\cM$~\eqref{mat:M} are actually:
\begin{equation}\label{mat:M2}
\cM_{a a}=-2\pi K(0)+iM\phi^{-1}(\la_a)\left(\phi_+(\la_a)-\phi_-(\la_a)\right).
\end{equation}

These functions are related by remarkable identities:
\begin{equation}
   \frac 1M\sul_{a=1}^N\frac{\sinh\zeta}
                            {\sin(\mu_k-\la_a\pm i\zeta)\sin(\mu_k-\la_a)}
                  \phi(\la_a)=\pm i \psi_\pm^{-1}(\mu_k),
\label{slavnovid}
\end{equation}
the detailed proof of which, based on identities for rational functions,
is given in Appendix A.

It is to mention that the factors $\phi(\la)$ and $\psi(\mu)$
were used earlier by N. Slavnov (we thank N. Slavnov for communicating  us
the results contained in his Ph. D. thesis).

\bigskip

Summing up the identities (\ref{slavnovid}) with plus and minus signs
one obtains the following equation:
\begin{equation}
   \frac{2\pi}{M}\sul_{a=1}^N K(\mu_k-\la_a)\phi(\la_a)
                   =i(\psi_+^{-1}(\mu_k)-\psi_-^{-1}(\mu_k)),
\label{phi->psi+-}
\end{equation}
which can be rewritten in an integral form by means of
proposition~\ref{period}:
\begin{equation}
   \int_{-\frac \pi 2}^{\frac \pi 2}K(\mu-\la)\phi(\la)\rho(\la)d\la=
       \frac i{2\pi}(\psi_+^{-1}(\mu)-\psi_-^{-1}(\mu))+O(M^{-\infty}).
\label{equation1}
\end{equation}
An analogous equation can be derived similarly
for the functions $\psi(\mu)$ and $\phi_\pm(\la)$:
\begin{equation}
    \int_{-\frac \pi 2}^{\frac \pi 2}K(\la-\mu)\psi(\mu)\rho(\mu)d\mu=
       \frac i{2\pi}(\phi_+^{-1}(\la)-\phi_-^{-1}(\la))+O(M^{-\infty}).
\label{equation2}
\end{equation}

The computation of the main order of
the functions $\phi_\pm(\la)$, $\psi_\pm(\mu)$ in the thermodynamic limit
is rather simple but
it is more complicated to prove that there are
no rational perturbative finite-size corrections.
We present this calculation in the Appendix C.
The result is
\begin{equation}
    \phi_\pm(\la)=\psi_\mp(\la)=\pm i+O(M^{-\infty}).\label{phipm}
\end{equation}

\bigskip

 Equation~\eqref{equation1} is used to calculate the function $\phi(\la)$.
By means of~\eqref{phipm}, it can be rewritten as
\[
\int_{-\frac \pi 2}^{\frac \pi 2}K(\mu-\la)\phi(\la)\rho(\la)d\la=
-\frac 1{\pi}+O(M^{-\infty}),
\]
which admits a unique solution
\[
\phi(\la)\rho(\la)=-\frac 1\pi+O(M^{-\infty}).
\]

\bigskip

Thus, one obtains the following expression for the diagonal
terms~\eqref{mat:M2} of the matrix $\cM$ in the thermodynamic limit:
\begin{equation}
  \cM_{a a}=2\pi\rho(\la_a)M-2\pi K(0)+O(M^{-\infty}).
\end{equation}
Hence the  elements of $\cM$ have the following form:
\begin{equation}
    \cM_{a b}=2\pi M\big\{\delta_{a b}\rho(\la_a)
                -\frac 1M  K(\la_a-\la_b)\big\}+O(M^{-\infty}).
\label{M}
\end{equation}

%%%%%%%%%%%%%%%%%%%%%%%%%%%%%%%%%%%%%%%%%%%%%%%%%%%%%%%%%%%%%%%%%%%%%%%%%%%%%%%%
\section{Fredholm determinant representation \label{sec:result}}
\setcounter{equation}{0}

In the thermodynamic limit, the determinants of
the matrices $\cN$ and $\cM-2\cV$
are replaced by Fredholm determinants:
\begin{align}
\label{norm}
   \det \cN &=(2\pi M)^N\biggl(\pl_{j=1}^N\rho(\la_j)\biggr)
              \big\{\det(\hat{I}+\hat{K})+O(M^{-\infty})\big\},\\
\label{i-k}
   \det (\cM-2\cV) &=(2\pi M)^N\biggl(\pl_{j=1}^N\rho(\la_j)\biggr)
               \big\{\det(\hat{I}-\hat{K}-2\hat{V})+O(M^{-\infty})\big\},
\end{align}
where $\hat{I}$ is the identity operator and $\hat{K}$ is the integral operator acting on the interval
$[-\frac \pi 2,\frac \pi 2]$
with the kernel $K(\la-\mu)$ (\ref{kernelK}). The kernel of the integral
operator $\hat{V}$ is
\begin{equation}
       V(\la,\mu)\equiv V(\la)=\frac{e^{2i\la}-\cosh\zeta}
                       {2\pi\sin(\la+i\frac \zeta 2)\sin(\la-i\frac \zeta 2)}.
\end{equation}

To calculate the coefficient in the representation~\eqref{turned} one has to
 use the proposition \ref{period'}, and the first order of the Taylor
developpement of $\delta\tilde{\la}_j$ which has been computed in
Appendix C:
\[\exp\bigg\{i\sul_{j=1}^N\delta\tilde{\la}_j\bigg\}=i+O(M^{-\infty}).\]

This leads to the following Fredholm determinant representation
for the function
$s_z(m)$~\eqref{szm}:
\begin{equation}
\label{FDR}
    s_z(m)=i(-1)^{m-1}\frac{\det(\hat{I}-\hat{K}-2\hat{V})}
                           {\det(\hat{I}+\hat{K})}+O(M^{-\infty}).
\end{equation}

The function
\[
\tilde{s}_z(m)=\frac {\bra{\Psi_2}\,\s_m^z\,\ket{\Psi_1}}{\l\Psi_1\ket{\Psi_1}}
\]
can be computed similarly, which gives
\[
\tilde{s}_z(m)=-s_z(m).
\]

Therefore the spontaneous staggered magnetization admits
the following representation in terms of Fredholm determinants:
\begin{equation}
\label{FDRSSM}
    s_0=\left|\frac{\det(\hat{I}-\hat{K}-2\hat{V})}
                   {\det(\hat{I}+\hat{K})}\right|+O(M^{-\infty}).
\end{equation}

\bigskip

What remains to do now is to compute explicitly these determinants in
order to obtain the Baxter formula. This can be done by means of Fourier
transformation.

As the kernel of the integral operator $\hat{K}$ depends only on the difference
of the two variables, and since we are in the regime $q>1$,
this operator can be diagonalized by Fourier transformation. Its eigenvalues
are thus obtained as the Fourier coefficients $k_n$ of the function $K(\la)$:
\begin{equation}
     k_n= \int_{-\frac \pi 2}^{\frac \pi 2}K(\la)e^{2i\la n}\,d\la
             =e^{-2\zeta |n|}=q^{-2|n|}.
\label{eigen}
\end{equation}
 Therefore the determinant of the operator $\hat{I}+\hat{K}$ is only
the infinite (convergent) product of its eigenvalues,
namely ($q>1$):
\begin{equation}
     \det(\hat{I}+\hat{K})=\pl_{n=-\infty}^{\infty}(1+q^{-2|n|})
                          =2\pl_{n=1}^{\infty}(1+q^{-2n})^2.
\end{equation}

After Fourier transformation, the operator $\hat{V}$ becomes
an infinite matrix which admits only one nonzero column $v_{n 0}$:
\begin{equation}
    v_{n 0}=\int_{-\frac \pi 2}^{\frac \pi 2}V(\la)e^{2i\la n}\,d\la.
\label{v}
\end{equation}
Since the operator $\hat{I}-\hat{K}$ is diagonalized by the
Fourier transformation,
only the diagonal element $v_{0 0}$
contributes to the determinant, namely
\begin{equation}
   \det(\hat{I}-\hat{K}-2\hat{V})=-2v_{0 0}\pl_{n=1}^{\infty}(1-q^{-2n})^2.
\end{equation}
The computation of the integral~\eqref{v} leads to $v_{0 0}=-1$.
Finally, from the expression~\eqref{FDRSSM} for the spontaneous magnetization,
one obtains the following result:
\begin{equation}
  s_0=\left(\pl_{n=1}^\infty\frac{1-q^{-2n}}{1+q^{-2n}}\right)^2+O(M^{-\infty}),
\label{Baxter1}
\end{equation}
which coincides with the Baxter formula.

%%%%%%%%%%%%%%%%%%%%%%%%%%%%%%%%%%%%%%%%%%%%%%%%%%%%%%%%%%%%%%%%%%%%%%%%%%%%%%%
% APPENDICES
%%%%%%%%%%%%%%%%%%%%%%%%%%%%%%%%%%%%%%%%%%%%%%%%%%%%%%%%%%%%%%%%%%%%%%%%%%%%%%%
\section*{Appendix A}
\renewcommand{\theequation}{A.\arabic{equation}}
\setcounter{equation}{0}

 In this appendix we give the proof of identity~\eqref{slavnovid}, which
provides a typical example of a proof for identities involving rational
functions. In particular, \eqref{mat:M} and~\eqref{mat:V} can be proved
similarly.

We thus consider the following identity
\begin{equation}
  \sul_{a=1}^N\frac{\sinh\zeta}{\sin(\mu_k-\la_a+ i\zeta)\sin(\mu_k-\la_a)}
  \frac{\pl_{m=1}^N\sin(\la_a-\mu_m)}{\pl_{m\neq a}\sin(\la_a-\la_m)}=
           i\pl_{m=1}^N
  \frac{\sin(\mu_k-\mu_m+ i\zeta)}{\sin(\mu_k-\la_m+ i\zeta)}.
\label{toprove}
\end{equation}
 To prove this relation it is convenient to introduce the exponential
variables:
\[v_k=e^{i\mu_k }, \quad u_a=e^{i\la_a}, \quad q=e^\zeta.\]
The left hand side of (\ref{toprove}) can be represented  using these variables as
\begin{equation}
  \mathrm{l.h.s.}=iv_k^2\pl_{m=1}^N\frac{u_m}{v_m}
         \sul_{a=1}^N\frac{q^2-1}{(v_k^2-u_a^2)(v_k^2-u_a^2q^2)}
  \frac{\pl_{m=1}^N (u_a^2-v_m^2)}{\pl_{m\neq a} (u_a^2-u_m^2)}.
\label{lhs}
\end{equation}
For the right hand side one obtains
\begin{equation}
  \mathrm{r.h.s.}=i\pl_{m=1}^N\frac{u_m(v_k^2-v_m^2q^2)}{v_m(v_k^2-u_m^2q^2)}.
\end{equation}
Consider the sum in (\ref{lhs}) as a function of $v_k^2$
\[W_1(v_k^2)=\sul_{a=1}^N\frac{q^2-1}{(v_k^2-u_a^2)(v_k^2-u_a^2q^2)}
\frac{\pl_{m=1}^N (u_a^2-v_m^2)}{\pl_{m\neq a} (u_a^2-u_m^2)},\]
which is to compare with the following function
\[W_2(v_k^2)=v_k^{-2}\pl_{m=1}^N\frac{v_k^2-v_m^2q^2}{v_k^2-u_m^2q^2}.\]
Both of them  are rational functions. When $v_k^2\rightarrow\infty$
one can easily see that $W_J(v_k^2)\rightarrow 0$, for $J=1,2$.
Also  both functions have only simple poles in the points $v_k^2=u_m^2q^2$
and their residues  in these points are
equal. Hence one can conclude that
\[W_1(v_k^2)=W_2(v_k^2),\]
which leads immediately to the identity (\ref{toprove}).

%%%%%%%%%%%%%%%%%%%%%%%%%%%%%%%%%%%%%%%%%%%%%%%%%%%%%%%%%%%%%%%%%%%%%%%%%%%%%%%
\section*{Appendix B}
\renewcommand{\theequation}{B.\arabic{equation}}
\setcounter{equation}{0}

We present here the proof of the propositions \ref{period} and \ref{period'}.

Let us begin with the proof of proposition \ref{period}.
Note first that, in case of variables $q_j,\ 1\le j \le N$ which are
homogeneously distributed inside an interval of lenght $\pi$ (i.e.
$q_{j+1}-q_j=\frac{\pi}N$), it is easy to prove  the
analogous   proposition (with $\rho(\lambda)=\frac1{2\pi}$):
\[\frac 1M \sul_{j=1}^N f(q_j)=
\frac 1{2\pi}\int_{-\frac \pi 2}^{\frac \pi 2}dq\,f(q)+O(M^{-\infty}).\]
We will reduce to this case by means of
the Bethe equations in the logarithmic form (\ref{bethelog}):
\begin{equation}
  p_0(\la_j)+\frac 1M\sul_{k=1}^N \theta(\la_j-\la_k)=\frac{2\pi n_j}{M}
\end{equation}
For a set of variables $\{\la_j\}_{1\le j \le N}$ parametrizing the ground
or the quasi-ground state, let us introduce the function
\[
  q_M(\la)=p_0(\la)+\frac1M \sum_{k=1}^N \theta (\la-\la_k),
\]
and the set of variables
$q_j=q_M(\la_j)=\frac{2\pi n_j}{M},\ 1\le j \le N$. These new variables
have a homogeneous distribution. Moreover, $q_M(\la)$ is an invertible
function for any $\Delta > 1$, at
least for $M > M_0 (\Delta )$ with $M_0$ large enough. For $\Delta >
2$, $M_0 (\Delta ) = 0$ (and does not depend on $\Delta$). It follows
from the estimate
$$
q_M'(\lambda )  = p_0 '(\lambda) -(2\pi /M)\sum_{k=1}^{N} K(\lambda -
\lambda _k ) \geq p_0 '(\pi /2) - \pi K(0) \nonumber
$$
(the r.h.s. is positive for $\Delta > 2$). For $1 \le \Delta \le 2$,
the existence of $M_0 (\Delta)$ follows from the fact that the solution
for the ground state does exist and is unique for any $M$
\cite{YY1,YY2}, and the density $\rho (\lambda) = (1/2\pi ) \lim _{M\to
\infty} q_M (\lambda)$ is positive in the thermodynamic limit.
Note at last that, since $q_M(\la +\pi)=q_M(\la)+\pi$, the function
$f\circ q_M^{-1}$ is $\pi$-periodic.
Hence
we can represent any sum on $j$ as an integral:
\begin{align}
  \frac 1M \sul_{j=1}^N f(\la_j)&=\frac 1M \sul_{j=1}^N f(q_M^{-1}(q_j)),\\
        &= \frac 1{2\pi}\int_{-\frac \pi 2}^{\frac \pi 2}dq\,f(q_M^{-1}(q))
             +O(M^{-\infty}),
\label{dqM}
\end{align}
which we can rewrite as an integral on $\la$:
\begin{equation}
  \frac 1M \sul_{j=1}^N f(\la_j)=
        \frac 1{2\pi}\int_{-\frac \pi 2}^{\frac \pi 2} d\la\,
      %\left(p_0'(\la)-\frac {2\pi}M\sul_{k=1}^N K(\la-\la_k)\right)
         q_M'(\la) f(\la)+O(M^{-\infty}).
\label{constr}
\end{equation}
One has thus to evaluate the derivative of $q_M$ in the thermodynamic limit.
We look for it in the following form:
\begin{equation}
    q_M'(\la)=p_0'(\la)-\frac{2\pi}M \sul_{k=1}^N K(\la-\la_k)
             =2\pi(\rho(\la)+\varphi_M(\la)),
\end{equation}
where the correction $\varphi_M$ is to be determined.
An integral equation for it can be obtained taking $f(\la)=K(\mu-\la)$:
from~\eqref{constr} one obtains
\[\int_{-\frac \pi 2}^{\frac \pi 2}
d\la\,K(\mu-\la)\varphi_M(\la)=-\varphi_M(\mu) + O(M^{-\infty}).\]
As all the eigenvalues~\eqref{eigen} of the integral operator
$\hat{K}$ are positive,
the only function satisfying this  constraint
is $0$ up to order $O(M^{-\infty})$.
Hence we proved that
\begin{equation}
   \frac 1M \sul_{j=1}^N f(\la_j)=\int_{-\frac \pi 2}^{\frac \pi 2}
       d\la\,\rho(\la)f(\la)+O(M^{-\infty}).
\label{qed}
\end{equation}

\bigskip

The proposition \ref{period'} can be derived directly from this proposition
and the analogous fact for the homogeneous
distribution of momenta. Really for any function $g(q)$ such that $g'(q)$ is
a periodic function one can easily prove
that
\[\frac 1M \sul_{j=1}^N g(q_j)=\frac 1{2\pi}\int_{-\frac \pi 2}^{\frac
\pi 2}dq\,g(q)+\frac {C_1}M+O(M^{-\infty}),\]
(as all further corrections are proportional to the integrals of second and
higher derivatives of $g(q)$ and these
integrals are evidently zero). Now one can continue the proof as for the
first proposition.
One should just mention that if $g'(\la)$ is a periodic function then
$\frac{dg(\la(q))}{dq}$ is also a periodic function.

%%%%%%%%%%%%%%%%%%%%%%%%%%%%%%%%%%%%%%%%%%%%%%%%%%%%%%%%%%%%%%%%%%%%%%%%%%%%%%%
\section*{Appendix C}
\renewcommand{\theequation}{C.\arabic{equation}}
\setcounter{equation}{0}

In this appendix, we present the calculation of the functions
$\phi_\pm(\la)$~\eqref{Phipm} and $\psi_\pm(\la)$ \eqref{Psipm}.

The functions  $\phi_\pm(\la)$ can be expressed in the exponential form,
\begin{equation}
   \phi^\pm(\la_a)=
       \exp\left\{
        \sul_{k=1}^N\left(\ln\sin(\la_a-\la_k-\delta\tilde{\la}_k\pm i\zeta)
        -\ln\sin(\la_a-\la_k\pm i\zeta)\right)
       \right\}.
\label{phiexp}
\end{equation}
One should note that if all $\delta\tilde{\la}_k$ are replaced by the
corresponding $\delta\lambda_k$, this expression is exactly equal to $-1$.
We will use
this property further to calculate the function  $\phi_\pm(\la)$ in the
thermodynamic limit.

Let us at first calculate the ``shift functions'' $\delta\tilde{\la}_k$
and
$\delta\lambda_k$.
The difference of two successive Bethe equations in the logarithmic
form~\eqref{bethelog} leads to:
\begin{equation}
   M\Big\{p_0(\la_j+\delta\lambda_j)-p_0(\la_j)\Big\}
   +\sul_{k=1}^N\Big\{\theta(\la_j+\delta\lambda_j-\la_k)
                                -\theta(\la_j-\la_k)\Big\}
   =2\pi.
\end{equation}
Representing the differences in the left hand side as Taylor series and
using proposition \ref{period} and Lieb
equation~\eqref{Lieb}, we obtain the following equation for $\delta\lambda$:
\begin{equation}
     M\sul_{k=1}^\infty \frac 1{k!}\rho^{(k-1)}(\la)(\delta\lambda)^k=1
          +O(M^{-\infty}).
\label{deltalambda}
\end{equation}
This equation defines uniquely (at order $O(M^{-\infty})$)
the shift function $\delta\lambda$ as a functional of $\rho(\la)$.
The solution of this
equation can be found as a series
on $\frac 1 M$:
\[\delta\la=
   \frac 1{\rho(\la)M}\left(1+\frac 1{2M}\left(\frac 1{\rho(\la)}\right)'
              +\frac 1{12M^2} \left(\frac 1{\rho^2(\la)}\right)''
              +O\left(\frac 1{M^3}\right)\right).\]
We will not use the explicit form of this solution in further calculations.

The calculation of the function $\delta\tilde{\la}_k$ is  more complicated.
Taking the difference between the Bethe
equations in the logarithmic form for the ground state and quasi ground
state one obtains:
\begin{equation}
   M\Big\{p_0(\la_j+\delta\tilde{\lambda}_j)-p_0(\la_j)\Big\}
   +\sul_{k=1}^N\Big\{\theta(\la_j+\delta\tilde{\lambda_j}
                      -\la_k-\delta\tilde{\lambda}_k)
                  -\theta(\la_j-\la_k)\Big\}=2\pi.
\end{equation}
Let us introduce the function $z(\la_k)=M\rho(\la_k)\delta\tilde{\lambda}_k$.
Using the proposition \ref{period} and Lieb
equation~\eqref{Lieb}, one obtains
the following equation for this function
\begin{equation}
    z(\la)+\int\limits_{-\frac \pi 2}^{\frac \pi 2}K(\la-\mu)z(\mu)d\mu=1
        +O\left(\frac 1 M \right).
\end{equation}
This equation admits a unique solution $z(\la)=\frac 12+O(M^{-1})$,
hence the expression for the first order of
 $\delta\tilde{\la}_k$:
\[\delta\tilde{\la}_k=\frac 1{2\rho(\la_k)M}+O\left(\frac 1{M^2}\right).\]
To obtain the next orders we compare the Bethe equations for
the ground and quasi-ground state, namely
the sums $\sum_{k=1}^N\theta(\la_j-\la_k)$
and  $\sum_{k=1}^N\theta(\la_j-\mu_k)$.  The function
$\theta(\la)$ is not periodic but its derivative is $\pi$-periodic,
and thus proposition \ref{period'} can be applied:
\[\frac 1M\left(\sul_{k=1}^N\theta(\la_j-\mu_k)-
\sul_{k=1}^N\theta(\la_j-\la_k)\right)=\frac{c}M+O(M^{-\infty})\]
where $c$ does not depend on $M$. As only the first order correction is
non-zero, we can use merely the first
order of $\delta\tilde{\la}_k$ to calculate $c$.
Finally one obtains
\[\sul_{k=1}^N\theta(\la_j-\mu_k)-\sul_{k=1}^N\theta(\la_j-\la_k)=
\pi+O(M^{-\infty}).\]
The Bethe equations for the ground and quasi-ground states can now
be written as
\begin{align}
  Mp_0(\mu_j)+\sul_{k=1}^N\theta(\mu_j-\mu_k) &=2\pi n_j,\\
  Mp_0(\la_j)+\sul_{k=1}^N\theta(\la_j-\mu_k) &=2\pi n_j-\pi+O(M^{-\infty}).
\end{align}
Taking the difference of these equations and using once again the Taylor
series and the Lieb equation
we obtain the equation defining $\delta\tilde{\la}$ as a functional of
$\rho(\la)$:
\begin{equation}
  M\sul_{k=1}^\infty \frac 1{k!}\rho^{(k-1)}(\la)(\delta\tilde{\lambda})^k
      =\frac 12+O(M^{-\infty}).
\end{equation}
This result is to be compared with~\eqref{deltalambda}. It follows that
\begin{equation}
  \delta\tilde{\lambda}[\rho(\la)]
  =\delta\lambda[2\rho(\la)]+O(M^{-\infty})\label{2rho}.
\end{equation}

To calculate the functions  $\phi_\pm(\la)$ we use our first remark:
replacing
$\delta\tilde{\la}_k$ by $\delta\lambda_k$ in the expression~\eqref{phiexp}
and developping it as a series on $\frac 1 M$ leads to
\begin{align*}
  -1 &=\exp\left\{\sul_{k=1}^N\Bigl(\ln\sin(\la_a-\la_k-\delta\la_k\pm i\zeta)
               -\ln\sin(\la_a-\la_k\pm i\zeta)\Bigr) \right\},\\
     &=\exp\left\{ -\sul_{k=1}^N\left(
      \frac{\cot(\la_a-\la_k\pm i\zeta)}{\rho(\la_k)M}\right)
      +\sul_{n=1}^\infty
      \frac{A_n[\rho(\la)]}{M^{n}}+O(M^{-\infty})\right\},
\end{align*}
where $A_n[\rho(\la)]$ are homogeneous functionals of $\rho(\la)$
which do not depend on $M$.
The first term of the development can be easily computed using
proposition~\ref{prop} once again:
\[\sul_{k=1}^N\left(\frac{\cot(\la_a-\la_k\pm i\zeta)}{\rho(\la_k)M}\right)=
\int_{-\frac \pi 2}^{\frac \pi 2}d\la\,\cot(\la_a-\la\pm i\zeta)
+O(M^{-\infty})=\mp i\pi+O(M^{-\infty}).\]
 Hence it follows that
\[A_n[\rho(\la)]=0.\]

Thus, due to the relation~\eqref{2rho}, the functions  $\phi_\pm(\la)$
can be represented as the following
series on $\frac 1 M$:
\[ \phi_\pm(\la)=\exp\left\{
  -\sul_{k=1}^N\left(\frac{\cot(\la_a-\la_k\pm i\zeta)}{2\rho(\la_k)M}\right)
  +\sul_{n=1}^\infty
    \frac{A_n[2\rho(\la)]}{M^{n}}+O(M^{-\infty})\right\}.\]
Since $A_n[\rho(\la)]$ are homogeneous functionals of $\rho(\la)$,
we obtain the very simple result:
\begin{equation}
  \phi_\pm(\la)=\pm i+O(M^{-\infty}).\label{phipm1}
\end{equation}
The functions   $\psi_\pm(\mu)$ can be computed similarly:
\begin{equation}
\psi_\pm(\mu)=\mp i+O(M^{-\infty})\label{psipm1}.
\end{equation}

%%%%%%%%%%%%%%%%%%%%%%%%%%%%%%%%%%%%%%%%%%%%%%%%%%%%%%%%%%%%%%%%%%%%%%%%%%%%%%%

\bibliographystyle{h-elsevier} %style de Nuclear Physics

\bibliography{biblio}
%%%%%%%%%%%%%%%%%%%%%%%%%%%%%%%%%%%%%%%%%%%%%%%%%%%%%%%%%%%%%%%%%%%%%%%%%%%%%%%

\end{document}